\documentclass{article}                     
\usepackage{graphicx}
\usepackage{amsmath,amsfonts,amsbsy,amssymb,fullpage,makeidx,array,graphicx,epsfig,enumerate}

\title{Editorial note to ``The beginning of the world from the point of view of quantum
theory''} \author{Jean-Pierre Luminet, Laboratoire Univers et Th\'eories\\ Observatoire de Paris-CNRS- Universit\'e Paris Diderot (France)\\email : jean-pierre.luminet@obspm.fr}

\begin{document}
\maketitle

\begin{abstract} This is an editorial note to accompany
reprinting as a Golden Oldie in the \emph{Journal of General
Relativity and Gravitation}\footnote{See
http://www.mth.uct.ac.za/~cwh/goldies.html.} of the famous note 
by Georges Lema\^{\i}tre on the quantum birth of the universe, published in \textit{Nature} in 1931\footnote{That republication will take place later this year in \textit{Gen. Rel. Grav.}}. We explain why this short  (457 words) article can be considered to be the true ``Charter'' of the modern Big Bang theory 
\end{abstract}

\bigskip

The year 1931 can undoubtedly be called
Georges Lema\^{\i}tre's \textit{annus mirabilis}. Indeed, major contributions to
relativistic cosmology by the Belgian physicist and priest appeared within a few
months:

\bigskip

a) \textit{A homogeneous universe of constant mass and increasing radius
accounting for the radial velocity of extra-galactic nebulae} [1] in the March 7
issue of the \textit{Monthly Notices of the Royal Astronomical Society}, as an
English translation of the article published four years earlier in French [2],
in which Lema\^{\i}tre was the first to interpret the astronomical data about
the galaxy redshifts by a positively curved space model in which the universe
slowly expanded from an equilibrium Einstein state at $t = -\infty$,

b) \textit{The expanding universe} [3] \footnote{Not to be confused with
\textit{L'Univers en expansion}, reproduced as a Golden Oldie as \textit{The
expanding Universe} [4].}, just following the previous one in the same
\textit{M.N.R.A.S.} issue, in which Lema\^{\i}tre calculated that the expansion
of space could be induced by a preceding phase of ``stagnation'' taking place
about $10^{10}$ years in the past,

c) The short note \textit{The beginning of the world from the point of view of
quantum theory}, published in the March 21 issue of \textit{Nature} and
reproduced here as a Golden Oldie\footnote{Due to its potential public impact and its non-technical character, the note was reprinted almost \textit{in extenso} in the \textit{New York Times} issue of May 19, 1931.},

d) \textit{Contribution to a discussion about ``The question of the relation of
the physical universe to life and mind''} [5], published in the October 24 issue
of \textit{Supplement to Nature}, in which Lema\^{\i}tre advocated an abrupt
beginning of the universe from an initial, superdense concentration of nuclear
matter called the ``primeval atom'',

e) \textit{L'expansion de l'espace} [6], a quantitative account of c) and d)
published in French\footnote{An English translation was published later in [7].}
in the November 20 issue of a Belgian scientific journal, where the author
developed his major cosmological ideas about the primeval atom hypothesis in an
extraordinary literary style,

\bigskip

\noindent and, since Lema\^{\i}tre was also fascinated by the brand new theory
of quantum mechanics, one should not forget to mention

f) \textit{L'ind\'etermination de la loi de
Coulomb}\footnote{\textit{Indeterminacy of Coulomb law.}} [8] in the August 8
issue of the \textit{Annales de la Soci\'et\'e Scientifique de Bruxelles}, in
which he applied Heisenberg's uncertainty principle to the Coulomb law, and

g) \textit{Sur l'interpr\'etation d'Eddington de l'\'equation de
Dirac}\footnote{\textit{On Eddington's interpretation of Dirac's equation.}} [9],
in the same issue, in which he investigated the mathematical structure of
quantum electrodynamics by using the formalism of quaternions.

\bigskip

In the middle of this string of pearls, the smallest (457 words) but the
brightest contribution c) can be considered to be the true ``Charter'' of the
modern Big Bang theory. To understand why, it must be explained and enlightened
by a larger \textit{corpus} of cosmological papers by Lema\^{\i}tre and other
leading cosmologists of the time during the crucial years 1930-1932. That is the
reason why the present editorial note, although devoted to a single-page
article, will take unusual proportions. It can also be seen as a celebration of
the 80th anniversary of Georges Lema\^{\i}tre's momentous ideas about the birth
and evolution of our universe.

\bigskip

{\bf \em Recession of galaxies and expanding universe}

\bigskip

Contrary to Friedmann (whose cosmological works were republished as a Golden
Oldie, see [10]), who came to astronomy only in 1921 -- 1922, that is to say
three years before his premature death only, Lema\^{\i}tre was closely related
to astronomy all his life. He always felt the absolute need for confronting the
observational facts and the general relativity theory (adding considerations
from quantum mechanics). He was, for example, much more aware than most of his
contemporaries of the experimental status of relativity theory, and that as
early as in his years of training [11]. Lema\^{\i}tre was no less a remarkable
mathematician, in the domain of fundamental mathematics (see his works on the
quaternions or St\"{o}rmer's problem) as well as in numerical analysis.

In short, the cosmological work of Lema\^{\i}tre was built in two phases.
Initially, he found independently of Friedmann that the Einstein field equations
of general relativity admitted non-static cosmological solutions. At the same
time, he took into account the observations on the recession velocity of
galaxies, to which he gave a physical meaning by interpreting them as an
experimental proof of an expanding space. In a next phase, Lema\^{\i}tre dared
an even more provocative assumption, which was however partly a logical
prolongation of the theory of the expanding universe: if the universe is today
expanding, in the past it was much smaller and denser; one remote day, it was
thus condensed into a ``primeval atom'', whose successive fragmentations due to
quantum processes made it such as it is now. Reviewed and improved during the
following decades, Lema\^{\i}tre's primeval atom hypothesis has become the
standard Big Bang model. Let us now follow in more details the evolution of
Lema\^{\i}tre's cosmological insights.

In his 1927 article \textit{Un univers homog\`ene de masse constante et de rayon
croissant, rendant compte de la vitesse radiale des n\'ebuleuses
extragalactiques} [2], Lema\^{\i}tre calculated the exact solutions of
Einstein's equations by assuming a positively curved space (with elliptic
topology), time varying matter density and pressure, and a non-zero cosmological
constant. He obtained a model with perpetual accelerated expansion, in which he
adjusted the value of the cosmological constant so that the radius of the
hyperspherical space $R(t)$ constantly increased from the radius of Einstein's
static hypersphere $R_E$  at $t = - \infty$. Therefore there was no past
singularity and no ``age problem''. The great novelty was that Lema\^{\i}tre
provided the first interpretation of cosmological redshifts in terms of space
expansion, instead of a real motion of galaxies: space was constantly expanding
and consequently increased the apparent separations between galaxies. This idea
proved to be one of the most significant discoveries of the century.

Using the available astronomical data, Lema\^{\i}tre provided the explicit
relation of proportionality between the apparent recession velocity and the
distance: ``Utilisant les 42 n\'ebuleuses extra-galactiques figurant dans les
listes de Hubble et de Str\"{o}mberg \footnote{ Lema\^{\i}tre combined the
redshifts published by  Str\"{o}mberg  -- who relied himself on redshifts
published earlier by Slipher  -- and Hubble' distances via magnitudes; in  his
book \textit{The Mathematical Theory of Relativity}, Eddington had also
published a redshift table, quoting data from Slipher who prepared that table
for him; see [12] for all references.}, et tenant compte de la vitesse propre du
Soleil, on trouve une distance moyenne de 0,95 millions de parsecs et une
vitesse radiale de 600 km/s, soit 625 km/s \`a $10^6$ parsecs. Nous adopterons
donc $R'/R = v/rc = 0,68 \times 10^{-27} {\rm cm}^{-1}$ (Eq. 24)''. Eq. 24 is
exactly what would be called later the Hubble law.

The significance of Lema\^{\i}tre's work remained unnoticed. Eddington, his
former PhD mentor to whom Lema\^{\i}tre had sent a copy, did not react. When
Lema\^{\i}tre met Einstein for the first time at the 1927 Solvay Conference, the
famous physicist made favorable technical remarks, but concluded by saying that
``from the physical point of view, that appeared completely abominable'' [13].
In 1929, Hubble [14] published new experimental data on the spectral redshifts
of extra-galactic nebulae, suggesting the linear velocity-distance relation $v =
Hr$ with $H = 600$ km/s/Mpc. This law was strictly identical to Lema\^{\i}tre's
Eq.24, with the same proportionality factor, but Hubble did not make the link
with expanding universe models. In fact Hubble never read Lema\^{\i}tre's paper;
he interpreted the galaxy redshifts as a pure Doppler effect (due to a proper
radial velocity of galaxies), instead of as an effect of space expansion.

A new opportunity for the recognition of Lema\^{\i}tre's model arose early in
1930. In January, in London, a discussion between Eddington and De Sitter took
place at a meeting of the Royal Astronomical Society. They did not know how to
interpret the data on the recession velocities of galaxies. Eddington suggested
that the problem could be due to the fact that only static models of the
universe were hitherto considered, and called for new searches in order to
explain the recession velocities in terms of dynamical space models. Having read
a report of the meeting of London [15], Lema\^{\i}tre understood that Eddington
and De Sitter posed a problem which he had solved three years earlier. He thus
wrote to Eddington to point out his communication of 1927 and requested him to
transmit a copy to de Sitter. This time, Eddington, who had not read the paper
at the right time, made apologies and reacted. He sent the note to de Sitter,
who answered very favorably in a letter to Lema\^{\i}tre, dated March 25,
1930.\footnote{Reproduced in [16], pp.104 -- 105.}

On his side, Eddington reorganized his communication to the following meeting of
the Royal Astronomical Society in May, to introduce Lema\^{\i}tre's ideas on
dynamical universes [17]. Then he published an important article [18] in which
he reexamined the Einstein static model and discovered that, like a pen balanced
on its point, it was unstable: any slight disturbance in the equilibrium would
start the increase of the radius of the hypersphere; then he adopted
Lema\^{\i}tre's model of the expanding universe -- which will be henceforward
referred to as the Eddington--Lema\^{\i}tre model -- and calculated that the
original size of the Einstein universe was about 1200 million light years, of
the same order of magnitude as that estimated by Lema\^{\i}tre. Interestingly
enough, Eddington also considered the possibility of an initial universe with a
mass $M$ greater or smaller than the mass $M_E$ of the Einstein model, but he
rejected the two solutions, arguing that, for $M > M_E$, ``it seems to require a
sudden and peculiar beginning of things'', whereas for $M < M_E$, ``the date of
the beginning of the universe is uncomfortably recent''.

Next, Eddington carried out an English translation of the 1927 Lema\^{\i}tre
article for publication in the \textit{Monthly Notices of the Royal Astronomical
Society} [1]. Here took place a curious episode: for an unexplained reason,
Eddington replaced the important paragraph quoted above (where Lema\^{\i}tre
gave the relation of proportionality between the recession velocity and the
distance) by a single sentence: ``From a discussion of available data, we adopt
$R'/R = 0,68 \times 10^{-27} {\rm cm}^{-1}$ (Eq. 24)''. Thus, due to Eddington's
(deliberate?) blunder, Lema\^{\i}tre will never be recognized on the same
footing as Edwin Hubble for being the discoverer of the expansion of the
universe.

Just following his translated article in the issue of \textit{M.N.R.A.S.},
Lema\^{\i}tre published a technical paper entitled \textit{The expanding
universe} [3], also communicated by Eddington, in which he studied the mechanism
of the initial expansion (see [19] for a detailed analysis). By dividing the
Einstein universe into cells in which matter condensed toward the centre of the
cell, Lema\^{\i}tre calculated that a diminution of the pressure on the edge of
the cells would induce a global expansion of space. He interpreted such a
diminution of the pressure as a diminution of the exchange of kinetic energy
between distant parts of the universe; in other words, the kinetic energy would
remain stagnant near the centre of the cells. Thus he introduced the phenomenon
of ``stagnation'' as the cause of the expansion of the universe: ``If, in a
universe of equilibrium, the pressure begins to vary, the radius of the universe
varies in the opposite sense. Therefore, stagnation processes induce
expansion''. Another original idea of this article was to generalize the
Birkhoff theorem of general relativity to describe the stagnation phenomenon
within the framework of a homogeneously expanding universe. Lema\^{\i}tre
visualized the initial Einstein universe as something like the dilute primeval
gas appearing in the Kant -- Laplace nebular hypothesis. At the end of his
paper, he considered the effect of a sudden stagnation process in which the
pressure dropped instantaneously to zero, and found that the epoch of the
rupture of equilibrium would have taken place some $10^{10}$ to $10^{11}$ years
ago, depending on the ratio between the pressure and the density.

\bigskip

{\bf \em Quantum birth of the universe}

\bigskip

Thus, at the beginning of 1931, the expansion of space appeared to be the only
coherent explanation to account for the astronomical observations. But the same
year when his vision of a dynamic universe was to be accepted by the scientific
community, including Eddington, de Sitter and Einstein, Lema\^{\i}tre dared to
make a much more outrageous assumption: if the universe is expanding now, must
it not have been much smaller and denser at some time in the past? Instead of
considering the static Einstein world as an initial stage from which the dynamic
model started, is it not more logical to think the universe as starting its
expansion from an extremely small and condensed state, governed by quantum
processes?

One of the reasons of this momentous idea was that, like many other physicists,
Lema\^{\i}tre was impressed by the new theory of quantum mechanics. Another
reason was to reply firmly to a communication delivered by Arthur Eddington at
the British Mathematical Association on January 5th, 1931 and published in the
March 9 issue of \textit{Nature} [20]. The British astrophysicist initially paid
tribute to Lema\^{\i}tre while declaring ``We recently learned, mainly thanks to
the work of Prof. Lema\^{\i}tre, that spherical space is expanding somewhat
fast''. Dealing with the role of entropy as an arrow of time, he considered
that, following time backwards, one would find more and more organization in the
world, up to a state of minimum entropy. But, for philosophical reasons,
Eddington refused to go back further in time up to the concept of singularity,
otherwise ``we have come to an abrupt end of space-time -- only we generally
call it the `beginning' ''. For him like for most others, this question laid
outside the range of science, and he added that ``philosophically, the notion of
a beginning of the present order of Nature is repugnant to me''.

In the Golden Oldie reproduced here, Lema\^{\i}tre argued that the world had
come into existence a finite time ago in an explosive event, which he likened to
a giant radioactive flash. Just like Eddington, he supposed that time and its
arrow are connected to the growth of the entropy. In the direction of increasing
time, the universe evolves to a state of infinite entropy, i.e. of complete
disorganization. In the direction of the past, the universe would have proceeded
from a state of zero entropy. Eddington had wondered whether the moment of zero
entropy could mark the beginning of the world, a concept that he had personal
reasons to discard. Lema\^{\i}tre disagreed and pointed out that entropy is a
measurement of proper time, and not of the time-coordinate; consequently,
Eddington was wrong to believe that the moment of minimal entropy separated
``before-creation'' from ``after-creation'' on an axis of universal time. It
should be seen, on the contrary, like an essential singularity, where the
concepts of space and time even lose their meaning. In order that spacetime can
exist within the framework of general relativity, one needs a tensor of
matter-energy, due to the identification of geometry and matter. The state of
matter with zero entropy constitutes a singularity of the matter-energy tensor
in the right-hand side of the field equations, which is equivalent to a
singularity in the curvature tensor in the left-hand side. There was no time nor
space prior to the state of condensation at zero entropy. It was the initial
singularity which created the space-time. Thus, the plurality and the diversity
of the physical world appeared to come from ``something'' physical, coinciding
with the $R = 0$ singularity of relativistic cosmological models. The
atom-universe exploded and plurality emerged. The entropy became nonzero, time
and its arrow also appeared.

The radical innovation introduced by Lema\^{\i}tre thus consisted in linking the
structure of the universe at large scales with the intimate nature of the atoms,
in other words in relating the early universe to quantum mechanics.\footnote{He
was not the first to suggest a connection between cosmology and quantum theory.
As early as 1925, Cornelius Lanczos introduced quantum mechanics in a
cosmological model, concluding that `The solutions of the quantum secrets are
hidden in the spatial and temporal closedness of the world' [21].} Lema\^{\i}tre
used the term ``single quantum'' and took care to stress that at this stage, the
laws of physics such as we know had no meaning anymore because the concepts of
space and time were not defined. It is the frontier of science such as
Lema\^{\i}tre conceived it, and in the present-day quantum cosmology nothing
clearly indicates that this frontier of physical knowledge, called the Planck
era, can be crossed.

Let us analyze in more detail Lema\^{\i}tre's argumentation. He begins by
stating that the number of distinct quanta is ever increasing in the course of
time.  He will explain better this assertion in the semi-popular paper published
later in the same year [6] (see also below). Let two volumes $V_1$ and $V_2$
contain heat radiation at temperatures $T_1$ and $T_2$, and let be $T$ the
equilibrium temperature of the total volume $V_1 + V_2$. From the law of energy
conservation it follows that $V_1 {T_1}^4 + V_2 {T_2}^4 = \left(V_1 + V_2\right)
T^4$, and the number of photons will increase proportionally to $\left(V_1 +
V_2\right) T^3 - V_1 {T_1}^3 - V_2 {T_2}^3$, which is always positive. The
demonstration is valid only for a gas of photons and not for material particles,
but Lema\^{\i}tre generalizes it by assuming in an intuitive way that the number
of particles sharing a given amount of energy is constantly increasing.

Next, Lema\^{\i}tre continues his short text by following closely an argument by
Niels Bohr published a few months before [22], according to which the concepts
of space and time in quantum mechanics have only statistical validity. As a
consequence, when the number of quanta was reduced to a single one, as assumed
to be the case at the beginning of the world, the notions of space and time
failed. They got meaning only when the original quantum began to disintegrate.
Therefore, the beginning of the world (namely the single quantum) happened ``a
little before'' the beginning of space and time. The phrasing is equivocal,
since ``a little before'' seems to imply a temporal sense, which would be
contradictory with the idea that time did not yet exist. Lema\^{\i}tre wanted to
say that space and time ermerged from the original quantum in a logical sense.

Now, what did he consider the original quantum to be made of? Lema\^{\i}tre
suggested that it might be a huge atomic nucleus, with an extremely large atomic
number corresponding to the total mass of the universe, and acting like a
quantum number. In 1931, nuclear physics was still in its infancy and the
neutron had not yet been discovered; but Lema\^{\i}tre knew about radioactive
processes, and he hypothesized that a huge atom would be unstable and
explosively decay into a large number of quanta. As he explained later, the word
``atom'' had to be taken in the Greek sense, as something completely
undifferentiated and deprived of physical properties.

In the final paragraph, Lema\^{\i}tre appealed to Heisenberg's uncertainty
principle to express the idea that the whole course of cosmic evolution was not
written down in the first quantum.

As underlined in [19], Lema\^{\i}tre's note was not an ordinary scientific
communication, but rather ``a visionary piece of cosmo-poetry that was meant to
open the eyes of the readers rather than convince them''. He wanted to make his
own view concerning the beginning of the world publicly known and understood by
everyone, thus he did not introduce any equation. In addition, he chose to sign
his communication as a private person, namely ``G. Lema\^{\i}tre, 40 rue Namur,
Louvain'', and not as a distinguished physicist and cosmologist, professor at
the University of Louvain.

It is time to recall that Lema\^{\i}tre was also a Catholic priest, and since
the creation of the universe a finite time ago is a dogma in Christian thought,
it might be tempting to jump to the conclusion that the explosive universe was
motivated by the aim to reconcile relativistic cosmology with religious belief.
It is interesting to point out that the manuscript (typed) version of
Lema\^{\i}tre's article, preserved in the Archives Lema\^{\i}tre at the
Universit\'e of Louvain, ended with a sentence crossed out by Lema\^{\i}tre
himself and which, therefore, was never published. Lema\^{\i}tre initially
intended to conclude his letter to \textit{Nature} by ``I think that every one
who believes in a supreme being supporting every being and every acting,
believes also that God is essentially hidden and may be glad to see how present
physics provides a veil hiding the creation''. This well reflected his deep
theological view of a hidden God, not to be found as the Creator in the
beginning of the universe. But before sending his paper to \textit{Nature},
Lema\^{\i}tre probably realized that such a reference to God could mislead the
readers and make them think that his hypothesis gave support to the Christian
notion of God.

\begin{figure}[h!]
\begin{center}
\includegraphics[scale=0.4]{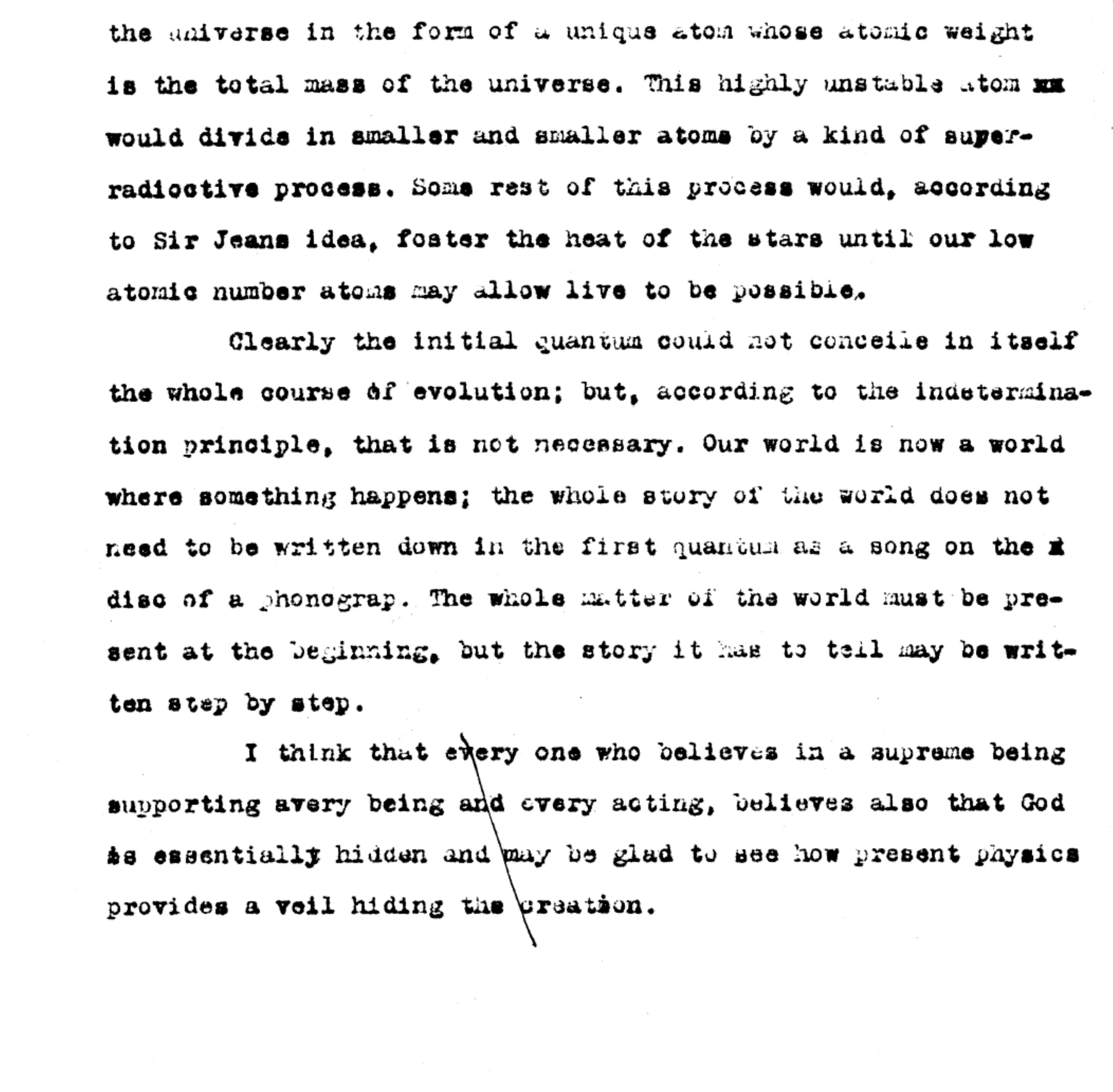}
\caption{A copy of the original Lema\^{\i}tre typescript, with the last paragraph crossed
by the author.}       
\end{center}
\end{figure}

As well analyzed in [23], Lema\^{\i}tre will preserve all his life the
conception of a supreme and inaccessible God, enabling him to keep the natural
origin of the world within the strict limits of physics, without mixing it with
a supernatural creation. As a priest just like a scholar in theology,
Lema\^{\i}tre was very conscious of the potential conflict -- or, on the
contrary, of the concordance -- between the Christian dogma of a world created
by God and the scientific theory of a universe formed approximately ten billion
years ago. However, Lema\^{\i}tre never confused science and religion. Contrary
to some other Christian cosmologists, he took care not to use one of these two
``ways of knowledge'' as a legitimisation of the other. He took, for example,
great care to distinguish between the ``beginning'' and the ``creation'' of the
world, and never spoke about the initial state of the universe in terms of
``creation'' (contrary to Friedmann, a fervent orthodox Christian, who
eventually appears more ``concordist'' than the Belgian priest). Lema\^{\i}tre
was convinced that science and theology dealt with two separate worlds.

\newpage

{\bf \em The Primeval Atom}

\bigskip

Lema\^{\i}tre had to convince himself that his model of an explosive universe
with finite age was physically realistic. He thus prepared a quantitative
article to be published in the fall of 1931. In the meantime, he accepted the
invitation of the British Association for Science to take part in its centenary
meeting, to be held in London on September 29, including a session on cosmology
devoted to ``The Question of the Relation of the Physical Universe to Life and
Mind''. Jeans, Eddington, Milne, de Sitter and Millikan made also scientific
contributions [24]. Among other questions, they had to deal with the problem of
the age of the universe, which, when deduced from the Hubble constant known at
the time, gave a value about 1.8 billion years, conspicuously smaller than the
time required for stellar evolution, as emphasised by Jeans and others.

Lema\^{\i}tre [5] argued (without any explicit calculation) that the problem
could be solved by making use of the stagnation process he had previously
introduced in the context of the Eddington -- Lema\^{\i}tre model [3]. But he
went much further by pushing forward his suggestion of an abrupt beginning of
the universe. As he said, ``a complete revision of our cosmological hypothesis
is necessary, the primary condition being the test of rapidity. We want a
'fireworks' theory of evolution [...] It is quite possible to have a variation
of the radius of the universe going on, expanding from zero to the actual
value.''

The singular creation of the universe had been briefly hypothesized by
Friedmann, but completely ignored by the scientific community. Lema\^{\i}tre
refined the argument and introduced for the first time (as far as we know) the
expression of \textit{primeval atom}: ``I would picture the evolution as
follows: at the origin, all the mass of the universe would exist in the form of
a unique atom; the radius of the universe, although not strictly zero, being
relatively small. The whole universe would be produced by the disintegration of
this primeval atom. It can be shown that the radius of space must increase.
[...] Whether this is wild imagination or physical hypotheses cannot be said at
present, but we may hope that the question will not wait too long to be
solved.''

Lema\^{\i}tre also suggested that the cosmic rays, which had been recently
discovered, were the fossils of the original explosion, as ``ashes and smoke of
bright but very rapid fireworks [...] We are led to the conclusion that the
stars were born some ten thousand million years ago without atmospheres, and
that the cosmic rays are outstanding features of the formation of a star''. The
origin of the cosmic rays was thought to be important evidence for the primeval
atom cosmology, but no trace of the idea has been found in Lema\^{\i}tre's
writings before this fall of 1931.

Eventually, the model of the primeval atom was quantitatively developed in
\textit{L'expansion de l'espace}, published in French in November 1931 [6].
Lema\^{\i}tre assumed a positively curved space (with elliptic topology),
time-varying matter density and pressure, and a cosmological constant such that,
starting from a singularity, the Universe first expanded, then passed through a
phase of ``stagnation'' during which its radius coasted that of the Einstein's
static solution, then started again in accelerated expansion.

The style of Lema\^{\i}tre contrasts drastically with that of Friedmann [10], in
the scientific argumentation as well as in the form. In the argumentation, the
approach of Friedmann was as axiomatic, as that of Lema\^{\i}tre -- himself a
remarkable mathematician -- was physical. As for the form, very literary
(adapted to that of the public conferences that Lema\^{\i}tre frequently gave),
it is a model of mixed rigor and lyricism, readable by almost everyone and which
testifies to the years of studies of Lema\^{\i}tre in graeco-latin humanities.
We reproduce below broad extracts of this extraordinary text in its English
translation [7], where the technical developments are omitted.

\bigskip

\begin{flushright}

\parbox{14cm}{
\renewcommand{\baselinestretch}{1.1}

{\small

\hspace{0.5cm} Following the Laplace and Kant cosmogonies, we became accustomed
to taking a very diffuse nebula as the starting point for evolution, a nebula
filling all space and becoming more and more condensed by splitting into partial
nebulae and finally into stars.

\hspace{0.5cm} This very old idea has been adapted to the recent progress of
astronomy. It has been recently expanded in that fine book [\textit{The Universe
Around Us}] which Sir James Jeans dedicated to the study of the universe. We are
now in a position to estimate the density of the primeval nebula by evaluating
the masses and the distances of less large condensations of stars called the
extra-galactic nebulae, which enclose all that we know of the universe. If the
actual mass of the stars was supposed to be distributed uniformly throughout the
whole space that they occupy, one would find that the primeval nebula must have
been more rarefied than the highest vacuum which our physicists can hope to
achieve in the laboratories. The density of the universe reduces to $10^{-30}$
gram per cubic centimeter, a figure which is generally considered to be reliable
within a factor of one hundred.

\hspace{0.5cm} The idea of a primeval nebula has to meet a very serious
difficulty which can be removed only with the help of the theory of relativity
and of non-Euclidian geometries: the different parts of the nebula are pulled
together by gravity, and it seems as though they should have to collapse toward
their center of gravity. A first element of solution is brought about by the
possibility that real space was not Euclidian but should obey the laws of
Riemann's elliptic geometry. Then there is no longer a center of gravity.

\hspace{0.5cm} [...]

\hspace{0.5cm} All points of the nebula remain uniformly distributed in space;
the distance between any two of them is always the same fraction of the total
length of the closest straight line on which they lie; but this length, equal to
$\pi R$, varies with the radius $R$; every distance varies in the same ratio as
the variation of the radius of space.

\hspace{0.5cm} To study in detail the variation of the radius of space, it is
necessary to appeal for the equations of general relativity. It is possible,
however, to illustrate the result of relativistic computations by elementary
considerations involving the laws of classical mechanics. This is possible
because laws of relativity are reduced to a limit to the laws of Newton, when
they are applied to an infinitely small volume.

\hspace{0.5cm} [...]

\hspace{0.5cm} These equations account for the dynamics of the universe; they
accustom us to thinking of the radius of the universe as a physical quantity,
able to vary. The manner in which these equations have been obtained must not be
regarded as a rigorous demonstration. A demonstration which is not open to
criticism can be deduced only from the general equations of relativity.
Nevertheless, the elementary considerations evolved hitherto may allow us, in
some degree, to grasp the physical significance of results involving more
abstract methods. Now we must explain what change must be made in these
equations, in order to account for the equilibrium of the Laplace nebula, and to
show how this change can be justified.

\medskip

\hspace{0.5cm} \textit{The Cosmological Constant}

\medskip

\hspace{0.5cm} One of the most important achievements of the theory of
relativity is the identification of the idea of mass with that of energy. Energy
is essentially a quantity which is defined, except for an additive constant;
mass, on the contrary, insofar as it affects the law of universal gravity, does
not involve any arbitrary constant.

}}

\begin{minipage}{14cm}{

{\small

\hspace{0.5cm} The identification of mass and energy, therefore, admits of a
choice of the constant of energy, or, inversely, of the introduction of an
arbitrary constant to the expression of the gravitational mass. The theory of
relativity teaches us the manner in which this arbitrary constant must be
introduced. The equations of gravity are obtained by integration of equations
which express both the conservation of energy and momentum. This integration
naturally introduces a constant of integration. But this constant of integration
is not added to the energy or to the total mass; it is added to the density. In
other words, the necessary adjustment between energy and gravitational mass is
made, not on the total mass, but on the density. This arbitrary constant, which
is introduced in the equations, has been called the ``cosmological constant'',
because it has no importance except in problems involving the whole universe.

\hspace{0.5cm} [...]

\hspace{0.5cm} The interpretation of the cosmological term is straightforward.
It means that an elastic force, which tends to increase the radius, is
superimposed on the Newtonian force, which tends to diminish it. A value of the
radius exists, called the equilibrium radius, for which these two forces
neutralize one another. The nebula of Laplace will last, provided that the value
of the radius be suitably adjusted to the value of the total mass of the nebula.

\hspace{0.5cm} Thus we have succeeded in making the Laplace nebula maintain
equilibrium. Let us not rejoice too soon, because we shall have to realize that
this equilibrium is quite precarious.

\hspace{0.5cm} [...]

\hspace{0.5cm} We can therefore conclude that the formation of local
condensations in the Laplace nebula in equilibrium must have upset this
equilibrium and initiated the universe.

\medskip

\hspace{0.5cm} \textit{Expansion of the Universe}

\medskip

\hspace{0.5cm} The hypothesis of Laplace has, therefore, as its consequence, the
expansion of space. Does this expansion take place, and with what speed it is
produced?

\hspace{0.5cm} In a space with increasing radius, the material points, the great
extra-galactic nebulae, for example, remain uniformly distributed in space.
Nevertheless, their mutual distances increase, all in the same ratio. Thus, if
we observe the extra-galactic nebulae, we shall be able to state that their
distances increase while remaining proportional to one another and therefore
that all extra-galactic nebulae have velocities of recession proportional to
their distance. The velocities of stars or of nebulae are observed through the
displacement of their spectral lines, known by the name of the Doppler -- Fizeau
effect. The spectrum of distant nebulae shows displacement toward the red,
corresponding to velocities of recession up to 10,000 kilometers per second; and
insofar as it is possible to judge their distances, these velocities are quite
proportional to this distance. Up to now, we possess about fifty measurements of
velocity,\footnotemark{}\footnotetext{$^{11}$Editor's footnote: In his 1929
article, Hubble displayed the data for 46 radial velocities; four of them were
negative, -- for the Andromeda galaxy M31, its two satellites M32 and NGC 205,
and for the Triangulum galaxy M 33 --, all the other were positive, from which
he deduced the velocity-distance relation of proportionality.} and, as a
consequence of all these measurements, we can estimate that a nebula, located at
a distance of one hundred million light-years (a distance at which it is still
possible to photograph the nebula) has a velocity of recession equal to
one-twentieth of the speed of light, that is, about 15,000 kilometers per
second.

\hspace{0.5cm} This result permits us to estimate the size which the nebula of
Laplace would have had originally, and it determines the initial radius of
equilibrium of space at about one billion light-years. The present value of the
radius depends on the estimate of the density of  matter. In utilizing the value
which we indicated at the beginning of this section, we find that it is equal to
a dozen times the initial radius.

\hspace{0.5cm} The present state of the expansion enables us to get some idea,
not only of the primeval nebula, but also of the epoch in which the local
condensations were formed while initiating the expansion of space.

\hspace{0.5cm} [...]

\hspace{0.5cm} One finds that, if the world began as a Laplace nebula in
equilibrium, the first general condensation of any importance which took place
in it, and which therefore initiated the expansion of space, could not have
occurred at an epoch dating back more than one hundred billion years.

\medskip

 }}

\end{minipage}

\begin{minipage}{14cm}{

{\small

\hspace{0.5cm} \textit{The Time-Scale}

\medskip

\hspace{0.5cm} To realize the importance of this result, one must not forget
that the cosmogony of Laplace-Jeans is a slow cosmogony. The primeval, gaseous
masses are condensed as a result of small inequalities in their initial
distribution and form the first condensations: the extra-galactic nebulae. As we
have just seen, this event dates back only hundred billion years, at a maximum.
These nebulae were still gaseous at that time. Weak condensations then formed,
by chance, and, as Jeans has shown, they must tend to increase provided that
they be of sufficient dimension, comparable to the mutual distances of the
stars. But how much time is necessary for these vast condensations to have the
opportunity to be formed and to be able to be concentrated in a sphere whose
diameter is a hundred thousand times smaller than their initial diameter?

\hspace{0.5cm} Jeans asks one hundred thousand billions years for this evolution
and I am not sure that he has proved whether this is enough; we can only give
him one- thousandth of this time.

\hspace{0.5cm} One hundred billion years is, at the most, fifty times the age
attributed to the earth. It is one hundred times the amount of time necessary
for the lunar tides to brake the rotation of our satellite and force it to turn
the same face constantly toward the Earth. It is only a thousand times the
amount of time which it takes light to come us from nebula which have been
photographed by our telescopes. Did evolution really take place according to
Laplace's theory, starting with extreme diffuseness and reaching the present
state of matter: stellar condensations dispersed in a virtual vacuum? Light
would not require one minute to cross the Sun, and it would need four years to
reach the nearest star. The stellar world, like the atomic world, seems to be
extraordinarily empty.

\hspace{0.5cm} A really complete cosmogony should explain atoms as suns, and
certainly atoms cannot have extreme diffuseness as their origin.

\medskip

\hspace{0.5cm} \textit{Radioactivity}

\medskip

\hspace{0.5cm} In the atomic field, we know about a spontaneous transformation
which can give us some idea of the direction of natural evolution; it is the
transformation of radioactive bodies. Disregarding photons and electrons whose
mass is nil or very small, an atom of uranium is ultimately transformed into an
atom of lead and eight atoms of helium. This is a transformation from a state of
greater condensation to one of lesser condensation. On the average, uranium can
remain extant only four or five billion years before making its transformation.
Thorium behaves in an analogous manner.

\hspace{0.5cm} If we had appeared on Earth one hundred billion years later,
there would have been no appreciable amounts of radioactive substances, and we
would doubtless have ended our table of elements at bismuth and lead. Does the
table of elements really end with
uranium?\footnotemark{}\footnotetext{$^{12}$Editor's footnote: As we know, the
answer is yes concerning the natural elements. The first element beyond uranium,
the neptunium -- atomic weight 93 -- was synthesised in 1940.} Have we not come
too late to know heavier elements which were almost completely disintegrated
before our birth? Are not radioactive transformations a faint residue of the
original evolution of the world and did they not take place, on the stellar
scale, several billion years ago?

\hspace{0.5cm} [...]

\hspace{0.5cm} Our universe bears the marks of
youth\footnotemark{}\footnotetext{$^{13}$Editor's footnote: A first reference to
a ``young'' universe is found in the famous poem of Lucretius \textit{De Natura
Rerum} (I$^{\rm rst}$ century BC), the kind of classical latin literature that
Lema\^{\i}tre had read.} and we can hope to reconstruct its story. The documents
at our disposal are not buried in the piles of bricks carved by the Babylonians;
our library does not risk being destroyed by fire; it is in space, admirably
empty, where light waves are preserved better than sound is conserved on the wax
of phonograph discs. The telescope is an instrument which looks far into space,
but it is, above all, an instrument which looks far into the past. The light of
nebulae tells us the history of hundred million years ago, and all the events in
the evolution of the world are at our disposal, written on fast waves in
internebular ether.

\hspace{0.5cm} [...]

}}

\end{minipage}

\begin{minipage}{14cm}{

{\small

\medskip

\hspace{0.5cm} \textit{The Primeval Atom}

\medskip

\hspace{0.5cm} The world has proceeded from the condensed to the diffuse. The
increase of entropy which characterizes the direction of evolution is the
progressive fragmentation of the energy which existed at the origin in a single
unit. The atom-world was broken into fragments, each fragment into still smaller
pieces. To simplify the matter, supposing that this fragmentation occurred in
equal pieces, two hundred sixty generations would have been needed to reach the
present pulverization of matter in our poor little atoms, almost too small to be
broken again.

\hspace{0.5cm} The evolution of the world can be compared to a display of
fireworks that has just ended: some few red wisps, ashes and smoke. Standing on
a well-chilled cinder, we see the slow fading of the suns, and we try to recall
the vanishing brilliance of the origin of the worlds.

\hspace{0.5cm} The sun-atom splinters into fragments held together by universal
attraction, fragments which splinter in their turn, hurling into the vacuum
particles which are fast enough to escape the attraction of the entirety, sparks
escaping from the burning crucible where the atom became a star. Rays travel in
a straight line in the still-increasing desert of space, until they encounter a
lost oasis, our galaxy, a chilled seed, our Earth, and discharge an
electrometer, proving the formation of the suns.

\hspace{0.5cm} Primeval nebula or primeval atom? Slow cosmogony or fast
cosmogony? Gaseous cosmogony or radioactive cosmogony? How far must the old
ideas be preserved? Was the Earth ejected in the atomic state by the sun-atom,
or was it separated from it in the gaseous phase? What are the properties of
giant atoms and the laws which govern their disintegration? It would be
premature to try to answer these questions.

\hspace{0.5cm} In concluding, we must indicate the manner in which the theory of
the expansion of the universe is adapted to the idea of the primeval atom. We
can conceive of space beginning with the primeval atom and the beginning of
space being marked by the beginning of time. The radius of space began at zero;
the first stages of the expansion consisted of a rapid expansion determined by
the mass of the initial atom, almost equal to the present mass of the universe.
If this mass is sufficient, and the estimates which we can make indicate that it
is indeed so, the initial expansion was able to permit the radius to exceed the
value of the equilibrium radius. The expansion thus took place in three phases;
a first period of rapid expansion in which the atom-universe was broken into
atom-stars, a period of slowing-up, followed by a third period of accelerated
expansion.\footnotemark{}\footnotetext{$^{14}$Editor's footnote: This is very
close to the time evolution of the so-called ``standard'' Big Bang model of
2011!} It is doubtless in the third period that we find ourselves today, and the
acceleration of space which has followed the period of slow expansion could well
be responsible for the separation of stars into extra-galactic nebulae.

\hspace{0.5cm} It is not completely proven that we are not in the first period
of expansion, and in this case, that the present expansion might not be capable
of making us exceed the equilibrium radius, which would therefore be quite
large. After having continued their movement of expansion for several billion
years, the nebulae would stop, then fall back toward one another, and finally
collide with one another, putting an end to the history of the world, with final
fireworks, after which the radius of space would again be reduced to zero.

\hspace{0.5cm} This hypothesis was proposed by Friedmann in 1922, and revived
recently by Einstein. Against it, there are the present estimates of density,
but these are not quite certain. Moreover, we can reassure ourselves by stating
that space is still extending and that, even if the world must finish in this
manner, we are living in a period that is closer to the beginning than to the
end of the world.

\hspace{0.5cm} But it is quite possible that the expansion has already passed
the equilibrium radius, and will not be followed by a contraction. In this case,
we need not expect anything sensational; the suns will become colder, the
nebulae will recede, the cinders and smoke of the original fireworks will cool
off and disperse.

 }}

\end{minipage}

\end{flushright}

As can be seen, both the style and the scientific contents were of an amazing
richness. Lema\^{\i}tre built his model from experimental data: the observation
of the redshifts of remote nebulae resulted from the expansion of space, but the
existence even of these nebulae imposed that, in its past, the universe
underwent local processes of condensation which gave them birth. For
Lema\^{\i}tre, the expansion of space and the condensation of matter were the
demonstrations of imbalances between two opposite cosmic forces: the gravity,
attractive, and the cosmological constant, repulsive. In addition, the
observational results constrained the evolution of the world to a short duration
and implied a fast cosmogony. According to Hubble measurements indeed, the
expansion rate was equal to 540 km/s/Mpc. With such a fast growth rate and
without a cosmological constant, the present universe should have some 2 billion
years of existence. However it was already known, by the study of radioactive
elements, that the age of the Earth was at least 4 billion years. Obviously the
Earth could not be older than the universe. Lema\^{\i}tre thus needed the
cosmological constant both to get an age of the universe compatible with that of
the Earth, and to leave enough time for galactic condensations to be formed.

Lema\^{\i}tre's model (cf. figure below) divided the evolution of the universe
into three distinct phases: two fast expansion phases separated by a period of
deceleration. The first phase was an expansion of explosive type, resulting from
radioactive decay of an atom-universe. The initial expansion was determined by
the mass of the primeval atom, ``almost equal to the present mass of the
universe''. The ``almost'' presumably referred to his early picture of the
primeval atom as a huge condensation of nuclei. It was known from nuclear
physics that an atomic nucleus is lighter than the sum of its constituent
particles by an amount known as the mass defect. Likewise, the primeval atom
would be somewhat lighter than the galaxies resulting from its explosion. For
this phase, Lema\^{\i}tre used the image of a ``fireworks'' which, if poetic, is
not less pedagogically debatable: it caused constant misunderstanding --
repeated by popular accounts -- presenting the beginning of the universe like an
explosion of matter localized in outer space.

The second phase of Lema\^{\i}tre's model corresponded to a quasi-equilibrium
between the density of matter and the cosmological constant, resulting in a
practically constant radius during a period of stagnation; the attractive
effects of gravitation being dominating at small scales, it was during this
phase that the density fluctuations were formed, which condensed later on to
give rise to the large scale structures of the universe, with stars grouped into
galaxies and galaxies into clusters. The formation of local condensations
disturbed the equilibrium conditions, which made the cosmological constant
predominant and started again the process of expansion. According to
Lema\^{\i}tre, the universe was presently in the third stage.

Technically, the solution was obtained starting from the relativistic equations
by supposing space with positive curvature and a cosmological constant $\lambda$
slightly higher than the Einsteinian value $\lambda_E = 1/{R_E}^2 = 2GM/ \pi c^2
{R_E}^3$, where $R_E$ was the equilibrium radius of the 1917 Einstein universe
model. As to the age of the universe, Lema\^{\i}tre mentioned as a possible
value ten billion years, but it could be considerably higher as it depended on
the value of the cosmological constant. The duration of the stagnation phase
depended essentially on $\lambda = \lambda_E (1 + \varepsilon)$, being
arbitrarily large when $\varepsilon$ tended to zero. For this reason, the
Lema\^{\i}tre's model was sometimes called ``hesitating universe''.

 \begin{figure}[h]
 \begin{center}
 \includegraphics[scale=0.5]{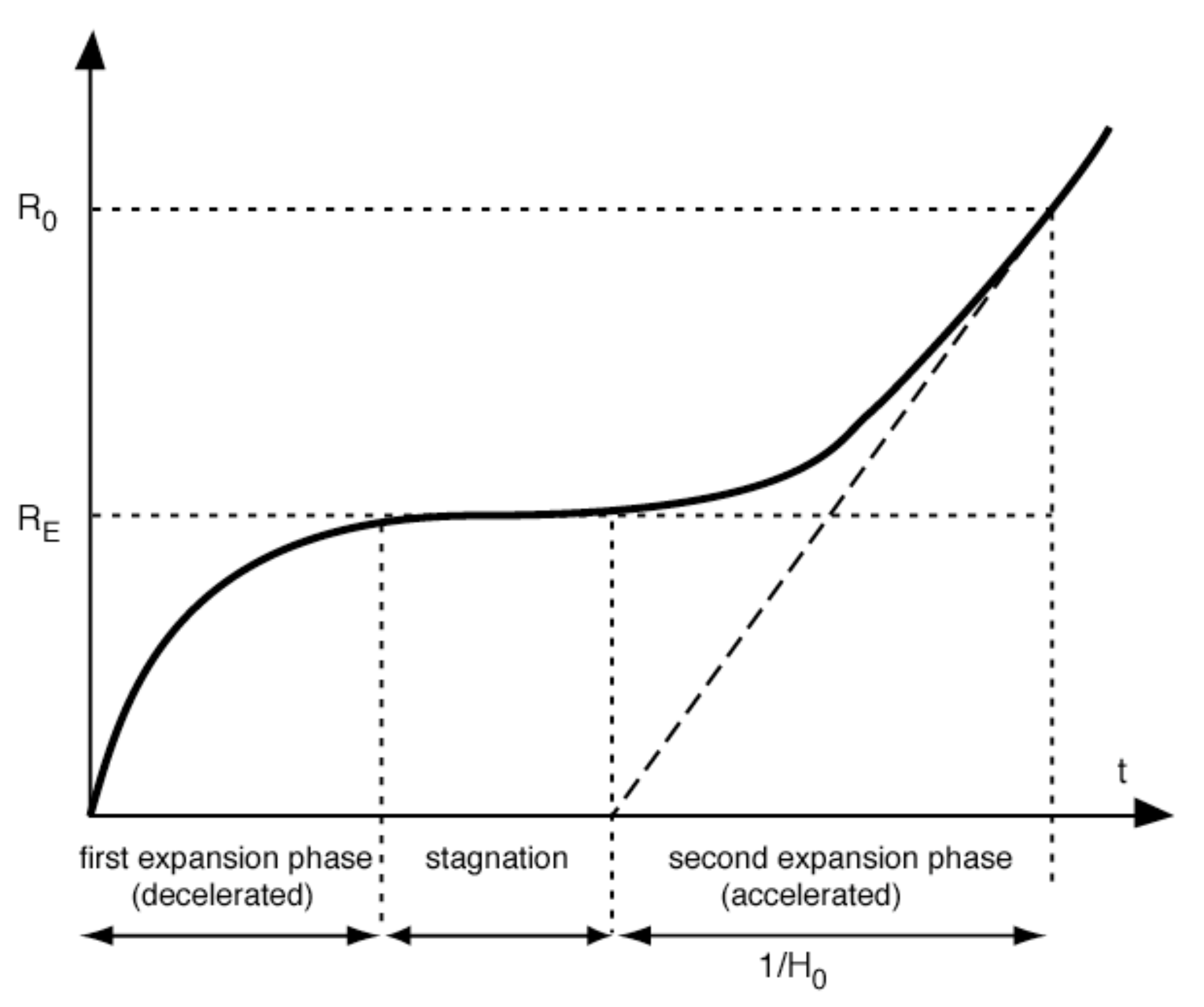}
 \caption{
Lema\^{\i}tre's ``hesitating universe''. The tangent to the expansion curve
measured today (oblique in dotted lines), i.e. the current expansion rate, gives
the Hubble time, often considered as an estimator of the age of the universe.
One sees clearly how the introduction of a cosmological constant and a
stagnation phase invalidates this estimate.
 }
 \end{center}
 \end{figure}

The reasoning of Lema\^{\i}tre was based on the will to use the new knowledge of
atomic physics and to link nebulae to the atoms, as he wrote it. Compared to his
model of 1927, which had a slow evolution, Lema\^{\i}tre proposed from now a
fast cosmology with an explosive origin, which, starting from the simplest,
generated the complex.

The Belgian physicist ended his paper with a brief discussion of the
eschatological aspects of his world model, arguing that in the far future, the
universe would inescapably end in heat death. Of course, all life would
irreversibly disappear...

\bigskip

{\bf \em Conclusion}

\bigskip

As a result of Lema\^{\i}tre's choice to publish his primeval-atom model in
French and in a semi-popular journal unknown to most physicists and astronomers,
it took some time until his article was noticed. When it became known, it was
poorly received by the majority. The fact that Lema\^{\i}tre was a mathematician
more than an astronomer, allied to his religious convictions, no doubt added to
a natural resistance towards cosmological revolutionary ideas.\footnote{One can
also wonder whether the literary quality of his work did not harm the
credibility of its scientific contents, in a community little accustomed to such
a flowery way of writing science. Still today, many scientists quickly and
pejoratively brand as ``popular'' a text raising the quality of the form to the
same level as that of the contents.} Eddington never accepted the primeval-atom
hypothesis or other ideas of the universe having an abrupt beginning a few
billion years ago. Like most other scientists, he felt uneasy about a created
universe, and this attitude was shared by the large majority of physicists and
astronomers in the 1930s. As we have mentioned above, this was an unfair
prejudice, because for Lema\^{\i}tre, as he expressed it several times, the
physical beginning of the world was quite different from the metaphysical notion
of creation, and for the priest-physicist, science and religion corresponded to
separate levels of understanding.

Therefore, even if Lema\^{\i}tre's hypothesis was mentioned in cosmological
reviews of the time by Tolman, de Sitter, Robertson and some others, it was not
assigned a physical reality. There were however a few exceptions. For instance,
the Harvard astronomer Donald Menzel wrote a quite enthusiastic article in a
popular science journal, beginning with ``Out of a single, bursting atom came
all the suns and planets of our universe! That is the sensational theory
advanced by the famous Abb\'e G. Lema\^{\i}tre, Belgian mathematician. It has
aroused the interest of astronomers throughout the world because, startling as
the hypothesis is, it explains many observed and puzzling facts.'' [25]. Also,
the quantum physicist Pascual Jordan supported Lema\^{\i}tre's model in a book
of 1936 [26]. The attitude of Einstein was less clear. As early as 1931 and
probably unaware of Lema\^{\i}tre's hypothesis, he derived from the Friedmann
equations a cyclic cosmological model in which the universe started expanding
from $R = 0$ and contracted into a ``big crunch'' [27], but he considered the
appearance of the singularity $R = 0$ to have no physical significance. Later
on, when he learnt about the primeval atom hypothesis, he considered it as
inspired by the Christian dogma of creation and totally unjustified from the
physical point of view. Einstein had also a great prejudice against the
cosmological constant he had originally introduced in his static model of 1917,
and that he considered as the ``greatest blunder'' of his life. It is probably
the reason why, in the new relativistic model he proposed in 1932 with de Sitter
[28]  -- a space with zero curvature and uniform density that expanded eternally
--, the term disappeared. Their model too belonged to the class of universes
with a singular beginning, so far that $R = 0$ for $t = 0$, but the authors did
not mention this feature, and did not even refer to either Friedmann or
Lema\^{\i}tre. After that, Einstein gave up research in cosmology.

Unfortunately, due to Einstein's authority, this over-simplified solution became
the ``standard model'' of cosmology for the next 60 years. However Lema\^{\i}tre
kept his original views. In 1933 he published another fundamental article about
cosmology, galaxy formation, gravitational collapse and singularities in the
\textit{Annales de la Soci\'et\'e Scientifique de Bruxelles}, translated to
English and reproduced as a \textit{Golden Oldie} more than ten years ago [4].
In that paper of 1933, Lema\^{\i}tre found a new solution of Einstein's
equations, known as the ``Lema\^{\i}tre -- Tolman'' or ``Lema\^{\i}tre -- Tolman
-- Bondi'' model, which is more and more frequently used today for considering
structure formation and evolution in the real Universe within the exact (i.e.
non perturbative) Einstein theory. In the less known \textit{Evolution of the
expanding universe} published in 1934 [29], he had a first intuition of a cosmic
background temperature at a few Kelvins: ``If all the atoms of the stars were
equally distributed through space there would be about one atom per cubic yard,
or the total energy would be that of an equilibrium radiation at the temperature
of liquid hydrogen.'' He also interpreted for the first time the cosmological
constant as vacuum energy: ``The theory of relativity suggests that, when we
identify gravitational mass and energy, we have to introduce a constant.
Everything happens as though the energy in vacuo would be different from zero.
In order that motion relative to vacuum may not be detected, we must associate a
pressure $p = - \rho c^2$ to the density of energy $\rho c^2$ of vacuum. This is
essentially the meaning of the cosmological constant $\lambda$ which corresponds
to a negative density of vacuum $\rho_0$ according to $\rho_0 = \lambda c^2/4
\pi G \sim 10^{-27}$ gr./cm$^3$''. Such a result would be rediscovered only in
1967 by Sakharov (the article has been republished as a \textit{Golden Oldie},
see [30]) and Zel'dovich [31] on the basis of quantum field theory; it is now
considered as one of the major solutions to the ``dark energy problem''.

\bigskip

To conclude, the following list summarizes the cosmological
questions\footnote{Not all of them were discussed here, for a complete survey
see [16].} discussed by Lema\^{\i}tre in the period 1927 -- 1934:

-- Expansion of space starting from an initial singularity

-- Dominating role of the cosmological constant in cosmic dynamics

-- Importance of the pressure of radiation in the early universe

-- Role of quantum theory at the origin of the universe

-- Problem of the age of the universe solved with a cosmological constant

-- Interpretation of the cosmological constant as the energy of the quantum
vacuum

-- Possibility of phoenix universes

-- Existence of relics of the early universe (cosmic residual temperature,
ultra-high energy cosmic rays)

-- Formation of galaxies due to random fluctuations of density

-- Topology of the universe.

\bigskip

On all these questions, Georges Lema\^{\i}tre showed an astonishing
perspicacity. This is the reason why the astronomer William McCrea, although an
adept of Milne's Newtonian cosmology, could declare in an article judiciously
entitled \textit{Some lessons for the future} [32]: ``Lema\^{\i}tre was a
scientist of superbly robust common sense. All of us who knew him must ever have
wished we had paid attention to his ideas. [...] Einstein, Eddington and Milne
may have been greater scientists than Lema\^{\i}tre, and more famous in their
day. But on the subject of cosmology and its importance for astronomy,
Lema\^{\i}tre had more to impart. He talked better sense.''

\bigskip

{\bf References}

\bigskip

\parindent=-0.5cm

\addtolength{\leftmargin}{0.5cm}

[1] G. Lema\^{\i}tre, \textit{A homogeneous universe of constant mass and
increasing radius accounting for the radial velocity of extra-galactic nebulae},
\textit{M.N.R.A.S.}, {\bf 41}, 483--490 (1931).

[2] G. Lema\^{\i}tre, \textit{Un univers homog\`ene de masse constante et de
rayon croissant, rendant compte de la vitesse radiale des n\'ebuleuses
extra-galactiques}, \textit{Annales de la Soci\'et\'e Scientifique de
Bruxelles}, s\'erie A, {\bf 47}, 49--59 (1927).

[3] G. Lema\^{\i}tre, \textit{The expanding universe}, \textit{M.N.R.A.S.}, {\bf
41}, 491--501 (1931).

[4] G. Lema\^{\i}tre, \textit{L'univers en expansion}, \textit{Annales de la
Soci\'et\'e Scientifique de Bruxelles} {\bf A53}, 51 (1933), reproduced as a
Golden Oldie as \textit{The expanding Universe} (translated by M. A. H.
MacCallum), \textit{Gen. Rel. Grav.} {\bf 29}, n$^\circ$5, 641 (1997). Editorial
note by A. Krasi\'nski, p. 637.

[5] G. Lema\^{\i}tre, Untitled contribution to a discussion on ``The question of
the relation of the physical universe to life and mind'', \textit{Supplement to
Nature}, {\bf 127}, October 24, 704--706 (1931).

[6] G. Lema\^{\i}tre, \textit{L'expansion de l'espace}, \textit{Revue des
Questions Scientifiques}, {\bf 20}, pp. 391--410 (1931).

[7] G. Lema\^{\i}tre, \textit{The primeval atom: An essay on cosmogony} (English
translation by B.H. and S.A. Korff), Van Nostrand Company, 1950.

[8] G. Lema\^{\i}tre, \textit{L'ind\'etermination de la loi de Coulomb},
\textit{Annales de la Soci\'et\'e Scientifique de Bruxelles}, s\'erie B, {\bf
51}, l$^{\mbox{\rm \`ere}}$ partie, pp. 12--16 (1931).

[9] G. Lema\^{\i}tre, \textit{Sur l'interpr\'etation d'Eddington de l'\'equation
de Dirac}, \textit{Annales de la Soci\'et\'e Scientifique de Bruxelles}, s\'erie
B, {\bf 51}, l$^{\mbox {\`ere}}$ partie, pp. 89--99 (1931).

[10] A. Friedmann, \textit{On the curvature of space} (translated by G. F. R.
Ellis and H. van Elst), \textit{Gen. Rel. Grav.} {\bf 31}, n$^\circ$ 12, 1991
(1999); (b) \textit{On the possibility of a world with constant negative
curvature of space} (translated by G. F. R. Ellis and H. van Elst), \textit{Gen.
Rel. Grav.} {\bf 31}, n$^\circ$12, 2001 (1999); Editorial note by A. Krasi\'nski
and G. F. R. Ellis, p. 1985; biography by A. Krasi\'nski, p. 1989; see also
Addendum in \textit{Gen. Rel. Grav.} {\bf 32}, 1937 (2000).

[11] G. Lema\^{\i}tre, \textit{La th\'eorie de la relativit\'e et
l'exp\'erience}, \textit{Revue des Questions Scientifiques}, 4$^{\rm e}$
s\'erie, vol. {\bf 9}, p. 346--374 (1926).

[12] G. Str\"{o}mberg, \textit{Analysis of Radial Velocities of Globular
Clusters and Non-Galactic Nebulae}, \textit{Astrophysical Journal}  {\bf 61},
353 (1925); V.M. Slipher, \textit{Nebulae}, \textit{Proceedings of the American
Philosophical Society}, {\bf 56}, 403 (1917); E. Hubble, \textit{Extra-Galactic
Nebulae}, \textit{Astrophysical Journal}  {\bf 64}, 321 (1926); A. Eddington,
\textit{The Mathematical Theory of Relativity}, Cambridge University Press
(1923), p.162 (reprinted 1960).

[13] G. Lema\^{\i}tre, \textit{Rencontres avec Einstein}, \textit{Revue des
Questions Scientifiques} t. 79 5e s\'erie, vol. {\bf 19}, p. 129--132 (1958).

[14] E. Hubble, \textit{A relation between distance and radial velocity among
extra-galactic nebulae}, \textit{Proceedings of the National Academy of
Sciences}  {\bf 15}, March 15, Number 3 (1929).

[15] A.S. Eddington, Remarks at the Meeting of the Royal Astronomical Society,
\textit{The Observatory}, vol. {\bf 53}, 39--40 (1930).

[16] J.-P. Luminet, \textit{L'invention du big bang}, \textit{Le Seuil}, Paris,
2004,

[17] A.S. Eddington, \textit{The Observatory}, vol. {\bf 53}, 162--164 (1930).

[18] A.S. Eddington, \textit{On the Instability of Einstein's Spherical World},
\textit{M.N.R.A.S.}  {\bf 90}, 668--678 (1930).

[19] H.S. Kragh, D. Lambert, \textit{The Context of Discovery: Lema\^{\i}tre and
the Origin of the Primeval-Atom Universe}, \textit{Annals of Science} {\bf 64},
445--470 (2007).

[20] A. Eddington,  \textit{The End of the World from the Standpoint of
Mathematical Physics}, \textit{Nature} {\bf 127}, 447--453 (1931).

[21] C. Lanczos, \textit{\"{U}ber eine zeitlich periodische Welt und eine neue
Behandlung des Problems der Atherstrahlung}, \textit{Zeitschrift f\"{u}r Physik}
{\bf 32}, 56--80 (1925).

[22] N. Bohr, \textit{The Use of the Concepts of Space and Time in Atomic
Theory}, \textit{Nature} {\bf 127}, 43 (1931).

[23] D. Lambert, \textit{Monseigneur Georges Lema\^{\i}tre et le d\'ebat entre
la cosmologie et la foi}, \textit{Revue th\'eologique de Louvain}, {\bf 28},
28--53 (1997); D. Lambert, \textit{L'itin\'eraire spirituel de Georges
Lema\^{\i}tre suivi de ``Univers et Atome'' (in\'edit de G. Lema\^{\i}tre)},
Bruxelles, Lessius, (2008).

[24] \textit{The Question of the Relation of the Physical Universe to Life and
Mind}, \textit{Supplement to Nature} {\bf 128}, 700--722 (1931).

[25] D. H. Menzel, \textit{Blast of Giant Atom Created Our Universe},
\textit{Popular Science Monthly}, 28--30 (Dec. 1932).

[26] P. Jordan, \textit{Die Physik des 20. Jahrhunderts}, Braunschweig, 1936;
English translation: \textit{Physics of the 20th Century}, New York, 1944.

[27] A. Einstein, \textit{Zum kosmologischen Problem der allgemeinen
Relativit\"{a}tstheorie}, \textit{Sitzungberichte der Preussischen Akademie der
Wissenschaften}, 235--237 (1931).

[28] A. Einstein and W. de Sitter, \textit{On the Relation Between the Expansion
and the Mean Density of the Universe}, \textit{Proceedings of the National
Academy of Sciences} {\bf 18}, 213--214 (1932).

[29] G. Lema\^{\i}tre, \textit{Evolution of the expanding universe},
\textit{Proceedings of the National Academy of Sciences} {\bf 20}, 12 (1934).

[30] A. D. Sakharov, \textit{Doklady Akademii Nauk SSSR} {\bf 177}, 70 (1967).
Golden oldie: \textit{Vacuum quantum fluctuations in curved space and the theory
of gravitation} (translated by Consultants Bureau), \textit{Gen. Rel. Grav.}
{\bf 32} n$^\circ$2, 365 (2000). Editorial note by H. J. Schmidt, p. 361.

[31] Y. B. Zel'dovich, \textit{Cosmological constant and elementary particles},
\textit{Soviet Physics JETP Letters} {\bf 6}, 316--317 (1967);  see also Y.B.
Zeldovich, \textit{The Cosmological Constant and the Theory of Elementary
Particles}, \textit{Soviet Physics Uspekhi} {\bf 11}, 381-393 (1968),
republished as a Golden Oldie in \textit{Gen. Rel. Grav.} {\bf 40} n$^\circ$7,
1562 (2008).

[32] W. McCrea, \textit{Personal recollections: some lessons for the future}, in
``Modern Cosmology in Retrospect'', Eds. B. Bertotti et al., Cambridge
University Press, p. 197--220 (1990).

\end{document}